\documentclass[final,5p,times,twocolumn]{elsarticle}

\usepackage{amssymb}
\usepackage{amsmath}

\journal{Solid State Sciences}

\def\kB{{k_{\rm B}}}

\begin{document}

\begin{frontmatter}



\title{First principles investigation of phase stability in the B-Pt alloy system}


\author{M. Widom} 


\address[]{Carnegie Mellon University,
            Department of Physics, 
            Pittsburgh,
            15213, 
            Pa,
            USA}

\begin{abstract}
  The B-Pt alloy system contains several Pt-rich phases exhibiting complex structures, many with partial site occupation. It also exhibits a deep (nearly 1000 degrees C) eutectic. We evaluate the {\em ab-initio} total energies of the crystalline solids to clarify the identity and character of the phases. Our work identifies inconsistencies in materials databases such as the Inorganic Crystallographic Structure Database and the ASM Phase Diagram Database, but our total energy calculations allow us to match up experimentally reported stable phases with specific structures. High temperature Gibbs free energy calculations in the liquid and solid states at the composition Pt$_2$B reveal that the depth of the eutectic arises from the low energy and high entropy of the liquid state.
\end{abstract}




\end{frontmatter}



\section{Introduction}
\label{sec:Intro}

The identities of phases in the boron-platinum system has been the subject of conflicting reports dating back to the 1950's~\cite{Buddery1951,Aronsson1959,Aronsson1960} and extending to the present day. Wald and Rosenberg~\cite{Wald1965} published a phase diagram showing three solid binary phases, Pt$_3$B$_2$, Pt$_2$B, and Pt$_3$B, along with their x-ray diffraction patterns but without definite structures. Matching phases identified in composition {\em vs.} temperature $(x,T)$ phase diagrams with structures as reported in crystallographic databases is an important task in general, and it is the primary goal for our present study of the B-Pt alloy system.

 A striking feature in the phase diagram is a eutectic of nearly 1000$^\circ$C deep. This eutectic has been utilized for the synthesis of elemental $\alpha-$boron~\cite{Horn1959}. As our second goal, we will extend our first principles study to include entropic effects at high temperature in order to explore the origin of the eutectic.

The assessed alloy phase diagram as reported by the ASM~\cite{Okamoto1990} is based primarily on the work of Wald and Rosenberg~\cite{Wald1965}. In this assessment, the phases were identified as having structure of Pearson type oS8 with space group {\it Cmcm} (phase name Pt$_3$B$_2$, formula PtB$_{0.67}$, Reference~\cite{Ellner1993}); Pearson type hP6 with space group {\it P}6$_3$/{\it mmc} and prototype anti-MoS$_2$ (phase name Pt$_2$B, formula Pt$_2$B, Reference~\cite{Ellner1993}); unknown cubic (phase name Pt$_3$B, formula Pt$_4$B, Reference~\cite{Hassler1979}). The Pearson type oS8 phase is shown as a line compound and is stable only above 600$^\circ$C. The other two have composition ranges at high temperatures but approach stoichiometric compositions in the limit of low temperatures.

The Inorganic Crystallographic Structure Database (ICSD) lists four alloy structures: Pearson type oS5 with space group {\it Cmcm} (formula Pt$_3$B$_{1.94}$, Reference~\cite{Ellner1993}); Pearson type mS18 with space group {\it C}12/{\it m}1 (formula Pt$_{12}$B$_{5.86}$, Reference~\cite{Sologub2015}); Pearson type hP4 with space group {\it P}6$_3$/{\it mmc} and prototype NiAs (formula PtB, Reference~\cite{Aronsson1959,Aronsson1960}); Pearson type hP6 with space group {\it P}6$_3$/{\it mmc} (formula Pt$_2$B, Reference~\cite{Hassler1979}). Presumably the phases with space group {\it Cmcm} match up, despite slightly conflicting Pearson types and formulas, and likewise the phases of space group {\it P}6$_3$/{\it mmc}. However, the ASM lists a phase of unknown cubic structure, and omits the structures of Pearson type mS18 and hP4.

First principles total energy calculations for all reported structures reveal their thermodynamic stability in the limit of low temperature, as well as their plausible stability at elevated temperatures. In several cases we explore stability over ranges of composition to account for partially occupied crystallographic sites. Once the structure of the Pt$_2$B solid phase is confirmed (as Pearson type mS18), we turn our attention to the origin of the deep eutectic. Through {\em ab-initio} molecular dynamics (AIMD) together with recently developed methods of entropy calculation~\cite{GaoWidom2018,e21020131,HuangKLiNa,Huang2022}, we estimate the temperature at which solid Pt$_2$B would melt congruently, and show that its low value is due to the low enthalpy and high entropy of the liquid.

\section{Methods}
\label{sec:Methods}

Electronic density functional theory~\cite{HohenbergKohn,KohnSham} allows us to calculate total energies of trial structures. All calculations are performed in the generalized gradient approximation~\cite{PBE} using the VASP code~\cite{Kresse96}. A plane-wave energy cutoff of 400 eV is applied consistently, and the $k$-point meshes are increased until energies are converged to within 1 meV/atom. We fully relax the structures at zero pressure to obtain their enthalpies, and we calculate enthalpies of formation,
\begin{equation}
  \label{eq:dH}
  H_{\rm For}(x)=H({\rm Pt}_x{\rm B}_{1-x})-x H({\rm Pt})-(1-x)H({\rm B}),
\end{equation} 
by subtracting the tie-line connecting the pure elements in their stable form~\cite{fezryb}, namely, the optimized structure of elemental boron~\cite{betaB2008}, and FCC platinum. Vertices of the convex hull of the set of enthalpies identify structures predicted to be stable at low temperature. For structures whose enthalpies lie above the convex hull, we define their instability energy $\Delta E$ as the enthalpy relative to the convex hull at the same composition~\cite{fezryb}.

Extensions to elevated temperature utilize ab-initio molecular dynamics (AIMD) simulations instead of relaxation. Our MD runs are performed in cubic cells of 480 atoms for the liquid, and 360-atom $1\times 5\times 4$ supercells of the Pt$_{12}$B$_6$.mS18 structure for the solid. Entropic contributions to the free energy are derived from positional correlation functions through recently developed methods as described in detail in section~\ref{sec:eutectic}. Additional details of our AIMD methods are discussed in~\ref{app:AIMD}.

\section{Results}
\label{sec:Results}

\subsection{Formation enthalpies at T=0K}
\label{sec:enthalpies}

In order to reconcile the phases and structures reported in the ASM and ICSD databases, we discuss the formation enthalpies of B-Pt structures in order of increasing Pt content, from elemental boron to elemental platinum. The ground state of elemental boron is a symmetry-broken variant of $\beta$-B (Pearson type hR141) as discussed in~\cite{betaB2008}. No B-rich compounds have been reported. The equiatomic PtB structure of Pearson type hP4~\cite{Aronsson1959,Aronsson1960} cannot be confirmed, as its positive formation enthalpy suggests it should not form even as a metastable state. However, this phase was also observed as an impurity phase in high pressure experiments~\cite{Whitney1963}, suggesting that it could possibly be stabilized either by impurities or perhaps at high pressure.

The structure reported by Wald at composition Pt$_3$B$_2$ was determined by Ellner and coworkers~\cite{Ellner1993} as centered orthorhombic with space group {\it Cmcm}. According to record number 67893 in the ICSD the platinum 4c site is 75\% occupied, and the boron 4a site is 50\% occupied, however the original paper~\cite{Ellner1993} refined the platinum site as fully occupied. Our calculations confirm the strong preference for full Pt occupation. In order to justify the nominal 3:2 stoichiometry (60~\% Pt), we considered a series of Pt$_4$B$_x$ structures by optimizing the placement of boron vacancies within a $2\times 2\times 2$ supercell of oS8. We find the enthalpy at 60\%~Pt lies significantly above the convex hull, but nearly touches it at 62.7-64.0\%~Pt. As this phase is reported to be stable at high temperatures only, we suggest that it is nonstoichiometric with a composition range extending above 60\%~Pt.

The convex hull exhibits a sharp minimum corresponding to the Pt$_{12}$B$_6$ structure of Pearson type oS18, confirming the stability of this phase and suggesting that this structure should be identified with the phase named Pt$_2$B. Stability of Pt$_2$B limits the possible composition range of the oS8 structure Pt$_4$B$_x$.

At higher Pt content, around 75\%~Pt, Sologub~{\em et al.}~\cite{Sologub2015} reported both low and high temperature variants of Pt$_3$B phases. The high T phase was identified as Pearson type hP6, previously reported by Hassler~{\em et al.}~\cite{Hassler1979} with an anti-MoS$_2$-[2H] structure, however, Hassler had identified the phase as Pt$_2$B. The low temperature phase was not identified, but its diffraction pattern shares several prominent peaks with the high temperature phase. We explored ordered patterns of boron vacancies within a $3\times 3\times 1$ supercell of hP6 and find their enthalpies lie close to the convex hull over a range of boron content, and actually touch the convex hull precisely at stoichiometry Pt$_3$B, suggesting the low T phase might be a modulated or vacancy-ordered supercell of hP6.

Hassler reported a cubic phase near concentration Pt$_4$B and presented a possible structure pictorially without giving a formal refinement. The structure appears similar to prototype WAl$_{12}$, but with metal vacancies at the 2a site, and a distribution of boron interstitials. We find this proposed structure to be unstable. Occupying the 2a site with Pt atoms reduces the instability without reaching the convex hull. However, the structure does become stable in B-Pt-X ternaries with X=Al or Cu placed on the 2a site, suggesting that the reported Pt$_4$B might have been stabilized by impurities.

\begin{figure}[t!]
\centering
\includegraphics[width=.49\textwidth]{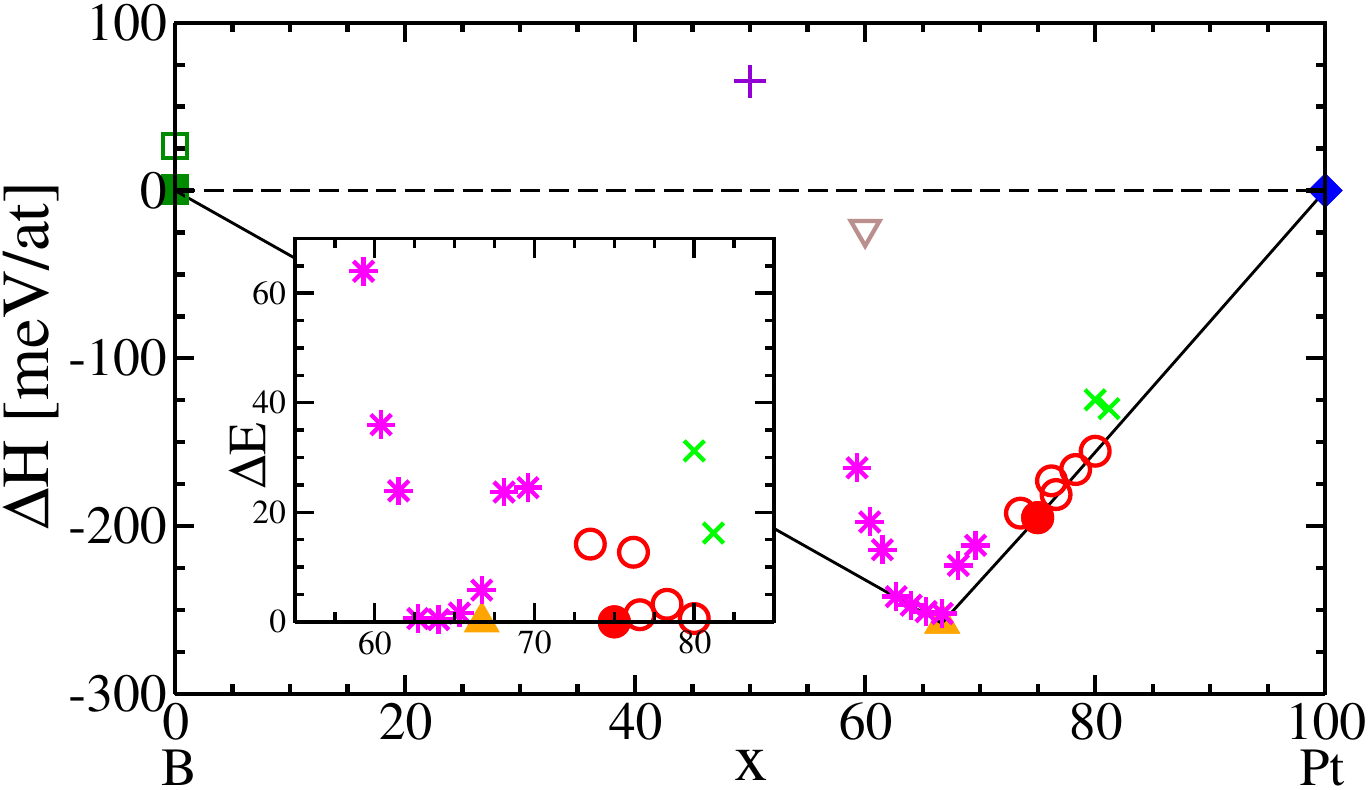}
\caption{\label{fig:data} Enthalpies of formation $\Delta H_{\rm For}$ (units meV/atom) of reported structures in the B-Pt alloy system. Plotting symbols: green squares (elemental boron~\cite{betaB2008}); purple plus sign (PtB.hP4~\cite{Aronsson1959}); brown down triangle (Pt$_3$B$_2$.oS8~\cite{Ellner1993}) magenta star (Pt$_4$B$_x$.oS8); orange up triangle (Pt$_{12}$B$_6$.mS18~\cite{Sologub2015}); red circle (hT-Pt$_3$B.hP6~\cite{Sologub2015}); green cross (Pt$_4$B.cI38~\cite{Hassler1979,Sologub2015}). See Table~\ref{tab:data} for numerical values. Fill symbols touch the convex hull. The inset plots instability energies $\Delta E$ over the composition range of interest. }
\end{figure}

\begin{table}[b!]
\centering
\begin{tabular}{ll|rrr}
  Series     & Pearson & $x$ & $\Delta H$ & $\Delta E$ \\
  \hline
  $\beta$-B  & aP282  &  0.0 & 0.0 & 0.0 \\
  $\beta$-B  & hR141  &  0.0 &  0.9 & 0.9 \\
  $\beta$-B  & hR105  &  0.0 &  26.6 & 26.6 \\
  $\alpha$-B & hR12   &  0.0 & 1.0 & 1.0 \\
  \hline
  PtB        &   hP4  & 50.0 &  65.1 & 258.5 \\
  \hline
  Pt$_3$B$_2$ &   oS8  & 60.0 & -23.1 & 209.0  \\
  \hline
  Pt$_4$B$_x$ &   oS8  & 59.3 & -165.3  & 64.0 \\
             &        & 60.0 & -197.6  & 36.0 \\
             &        & 61.5 & -214.2  & 23.9 \\
             &        & 62.7 & -242.1  &  0.6 \\
             &        & 64.0 & -247.2  &  0.4 \\
             &        & 65.3 & -251.1  &  1.6 \\
             &        & 66.7 & -252.2  &  5.8 \\
             &        & 68.1 & -223.5  & 23.7 \\
             &        & 69.6 & -211.6  & 24.5 \\
  \hline
  B$_6$Pt$_{12}$ &   mS18 &  66.7 & -257.9 & 0.0 \\
  \hline
  hT-Pt3B    &  hP6    & 73.5 & -192.3  & 14.2 \\
             &         & 75.0 & -195.0  & 0    \\
             &         & 76.2 & -173.1  & 12.7 \\
             &         & 76.6 & -181.3  & 1.3  \\
             &         & 78.3 & -166.4  & 3.2  \\
             &         & 80.0 & -155.4  & 0.6  \\
  \hline
  Pt$_4$B    &      cI38 & 80.0 & -124.8 & 31.2 \\
             &          & 81.2 & -130.0 & 16.2 \\
  \hline
  Pt         & cF4     & 100.0 & 0.0 & 0.0
\end{tabular}
\caption{\label{tab:data}Formation enthalpies $\Delta H$ and instability energies $\Delta E$ as plotted in Fig.~\ref{fig:data}, in units of meV/atom.}
\end{table}


\subsection{Eutectic transformation}
\label{sec:eutectic}

Prediction of eutectic transformations~\cite{HuangKLiNa} requires separate free energy models for the liquid phase and all competing solid phases across the composition range. Here we attempt a simpler study, seeking the exchange of thermodynamic stability between the low temperature solid Pt$_{12}$B$_6$ phase and the high temperature liquid of composition Pt$_2$B. This would represent congruent melting, and inspection of the eutectic region of the ASM phase diagram reveals the transformation is indeed nearly congruent at 67\%~Pt.

Our ab-initio calculation of the temperature-dependent Gibbs free energy utilizes ab-initio molecular dynamics (AIMD) to capture the thermal expansion and temperature-dependent internal energy.  The NPT ensemble is applied initially to obtain thermal expansion, followed by NVT simulations for energy and entropy calculation. Our NPT and NVT runs are accelerated using the machine learning features of VASP~\cite{VASP-ML}. Further details of the AIMD runs are discussed in \ref{app:AIMD}.

\begin{figure}[t!]
\centering
\includegraphics[width=.49\textwidth]{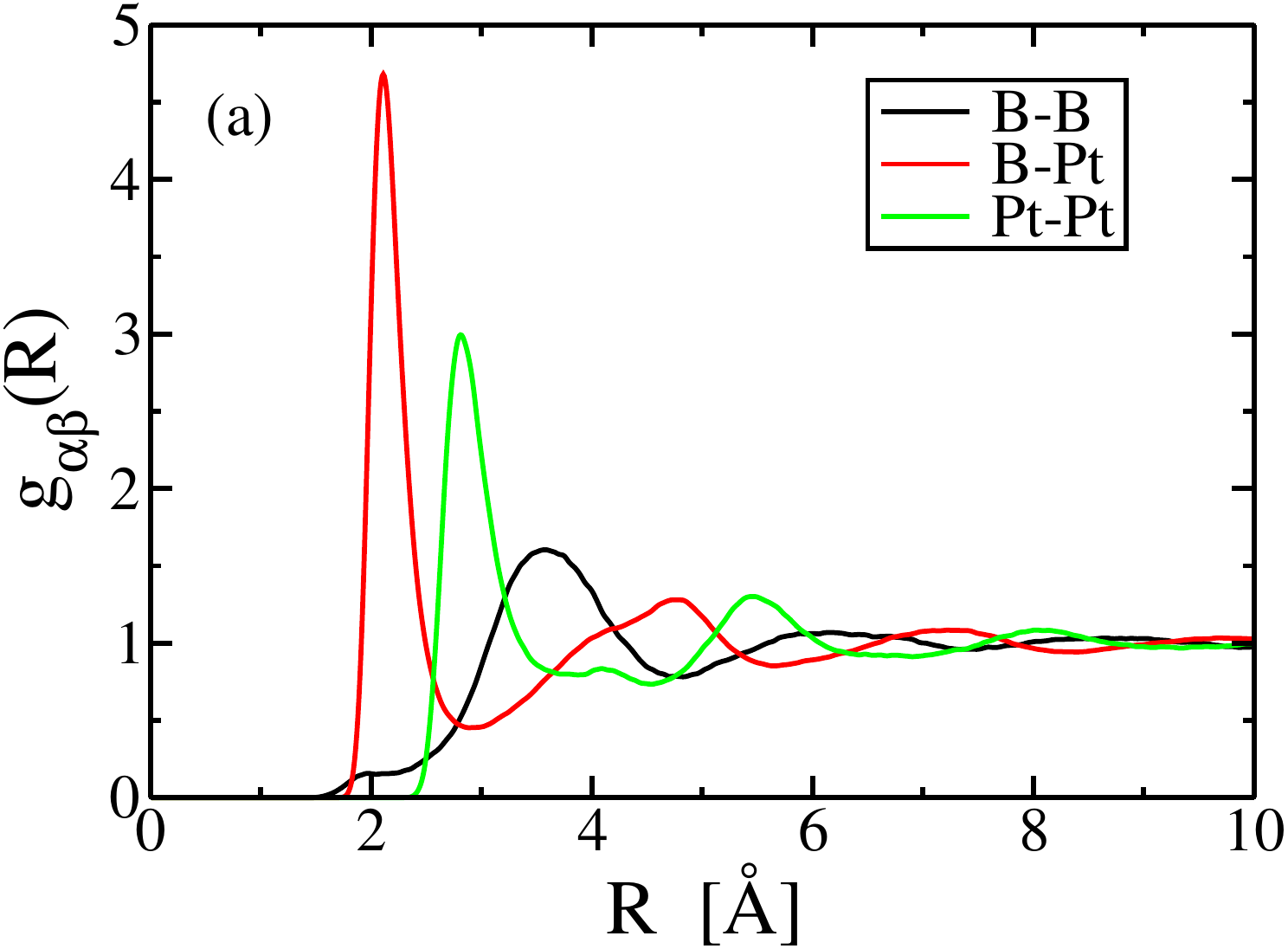}
\includegraphics[width=.49\textwidth]{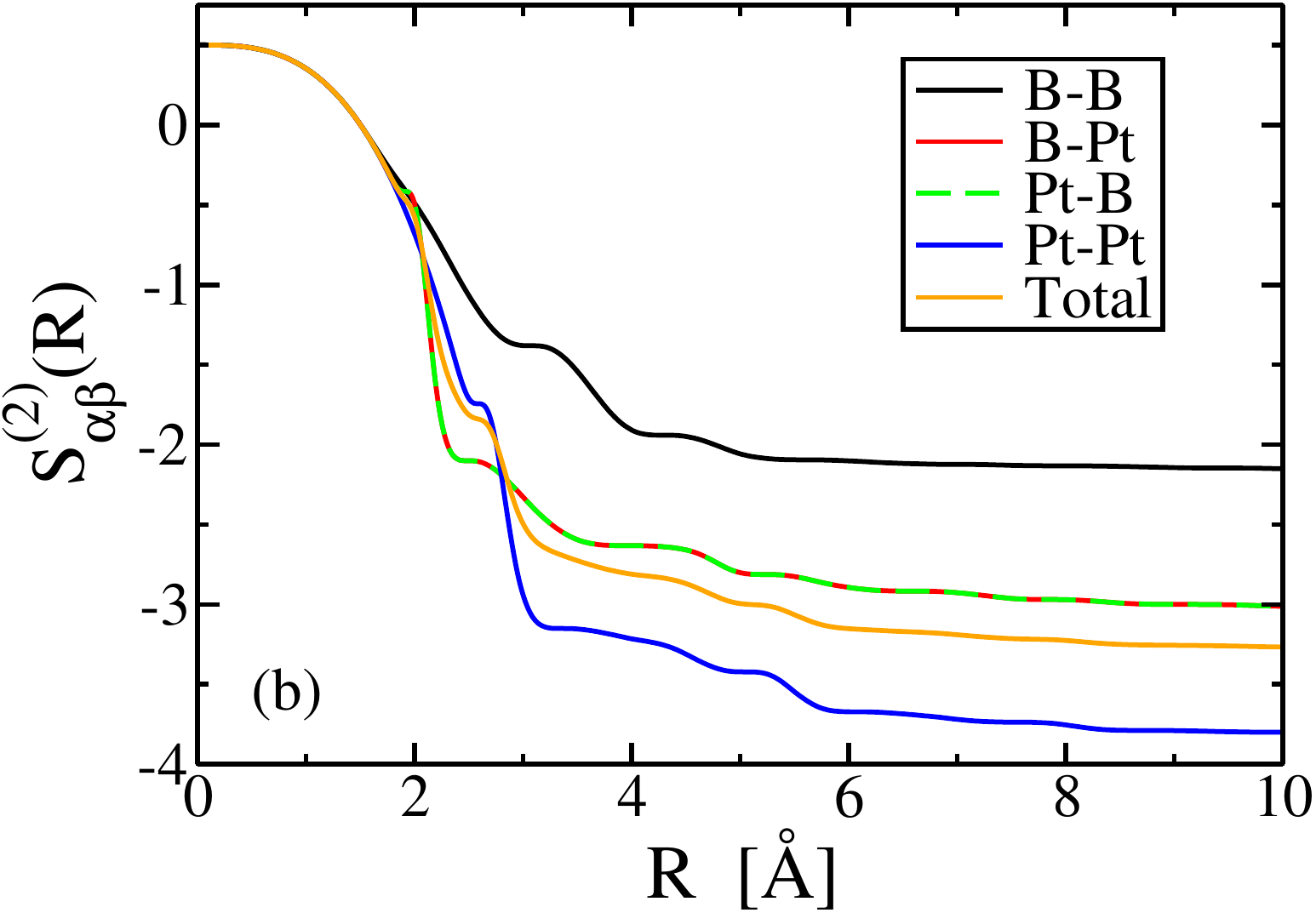}
\caption{\label{fig:gab} (a) Radial distribution functions at T=1173K. (b) convergence of integrals with respect to cutoff $R$, separated according species pairs $\alpha\beta$, and their composition-weighted sum, $S_2 = \sum_{\alpha\beta} x_\alpha x_\beta S^{(2)}_{\alpha\beta}$.}
\end{figure}

Liquid state entropy is derived from the simulated radial distribution functions~\cite{GaoWidom2018,e21020131,HuangKLiNa}, $g_{\alpha\beta}(r)$. We start from the single-particle entropy (in units of $k_{\rm B}$ per atom)
\begin{equation}
  \label{eq:S1}
  S_1 = \frac{3}{2}-\sum_\alpha\ln{(\rho x_\alpha\lambda_\alpha^3)}
\end{equation}
where $\rho$ is the atomic density, $x_\alpha$ is the mole fraction of chemical species $\alpha$ and $\lambda_\alpha=\sqrt{h^2/2\pi m_\alpha k_B T}$ is the thermal de~Broglie wavelength. We add to $S_1$ a two-body correction term, $S_2=S_{\rm Fluct}+S_{\rm Info}$, where the fluctuation term
\begin{equation}
  \label{eq:sfluct}
  S_{\rm Fluct} = \frac{1}{2} \sum_{\alpha,\beta}x_\alpha x_\beta \left(1 + \rho \int_0^R dr\, 4\pi r^2~ (g_{\alpha\beta}(r)-1)\right)
\end{equation}
is positive but very small (it is proportional to the isothermal compressibility) and the information term
\begin{equation}
  \label{eq:sinfo}
  S_{\rm Info} = -\frac{1}{2}\rho \sum_{\alpha,\beta}x_\alpha x_\beta \int_0^R dr\, 4\pi r^2~ g_{\alpha\beta}(r)\ln g_{\alpha\beta}(r)
\end{equation} is negative-definite and reflects the entropy reduction due to the information content of the radial distribution functions. Fig.~\ref{fig:gab}a illustrates the simulated radial distribution functions at T=1173K. Note the near absence of B-B neighbors that would occur around 2~\AA, and the strong peak for B-Pt neighbors. Fig.~\ref{fig:gab}b illustrates the convergence of the integrals in Eqs.~\ref{eq:sfluct} and~\ref{eq:sinfo} with respect to the upper limit of integration $R$. We adopt the value at $R=10$~\AA~ as our converged entropy loss due to spatial correlations.

Solid state entropy is based on the covariance of atomic displacement method (CAD~\cite{Huang2022,PhuthiCAD}). Each atom $i$ oscillates around its mean position with instantaneous displacement ${\bf u}_i$. We extract a covariance matrix $\Sigma_{\cal U}$ for the complete set of all $N$ atomic displacements ${\cal U} = \{{\bf u}_i\}$ by averaging over the full simulation trajectory, followed by symmetrizing using the rotations of the crystallographic space group and the translations of the simulated supercell. The entropy
\begin{equation}
  \label{eq:S_CAD}
  S = \frac{1}{2}\ln(\det{\Sigma_{\cal U}}) +
  \frac{3}{2}\sum_{i=1}^N\ln{(m_i k_B T/h^2)} + \frac{3}{2}N
\end{equation}
can be considered as a quasiharmonic entropy arising from a temperature-dependent effective potential. Sampling errors for simulation time $t$ reduce the calculated entropy below its actual value by an amount of order the inverse simulation time, which allows us to extrapolate to the long time limit~\cite{3body}.

\begin{figure}[htb]
\centering
\includegraphics[width=.49\textwidth]{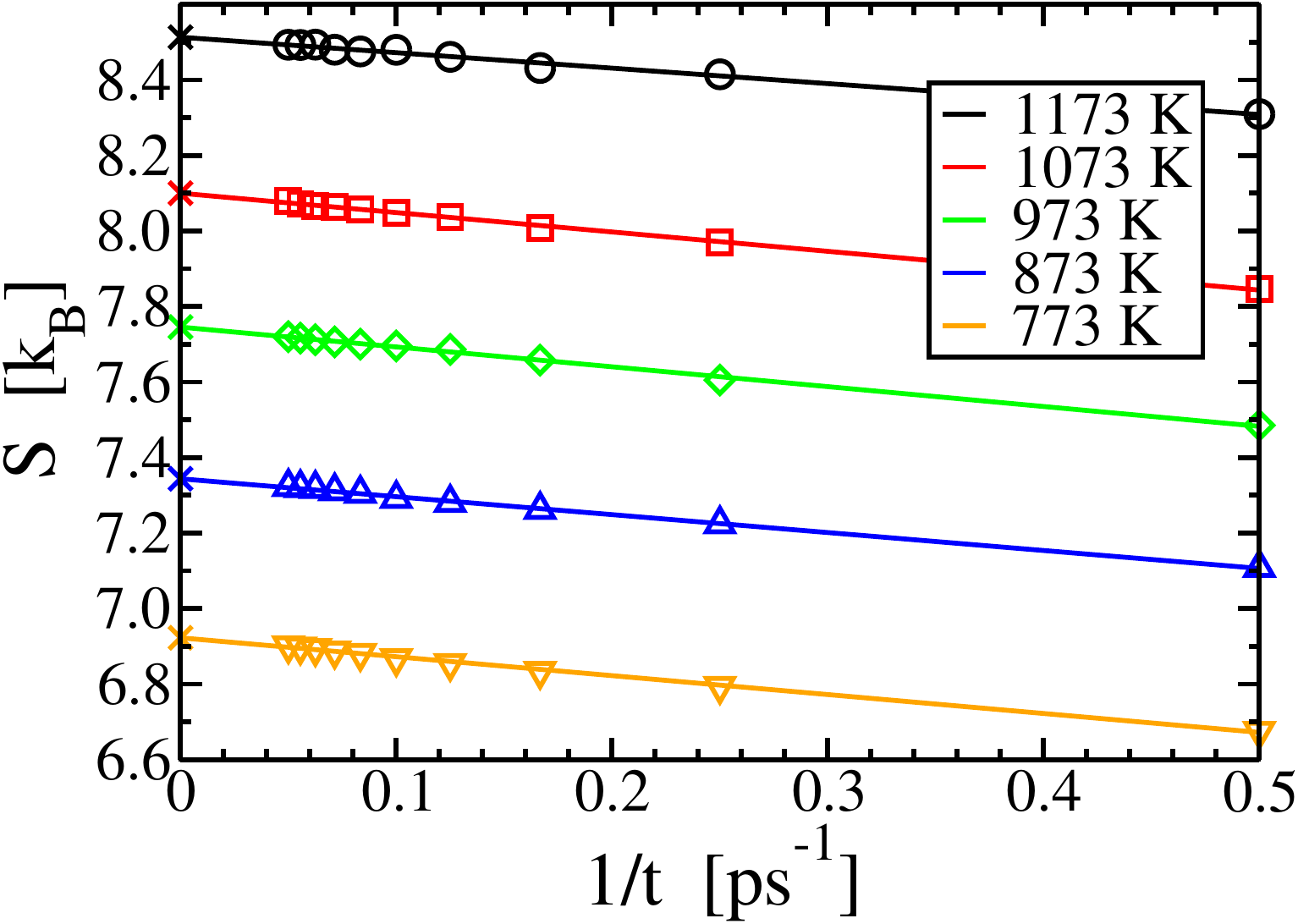}
\caption{\label{fig:Svst} Entropy of Pt$_{12}$B$_6$.mS18 calculated through the covariance of atomic displacements, plotted as a function of inverse simulation time. Extrapolated values for $1/t\to\infty$ are marked with $X$.}
\end{figure}

\begin{figure}[t!]
\centering
\includegraphics[width=.49\textwidth]{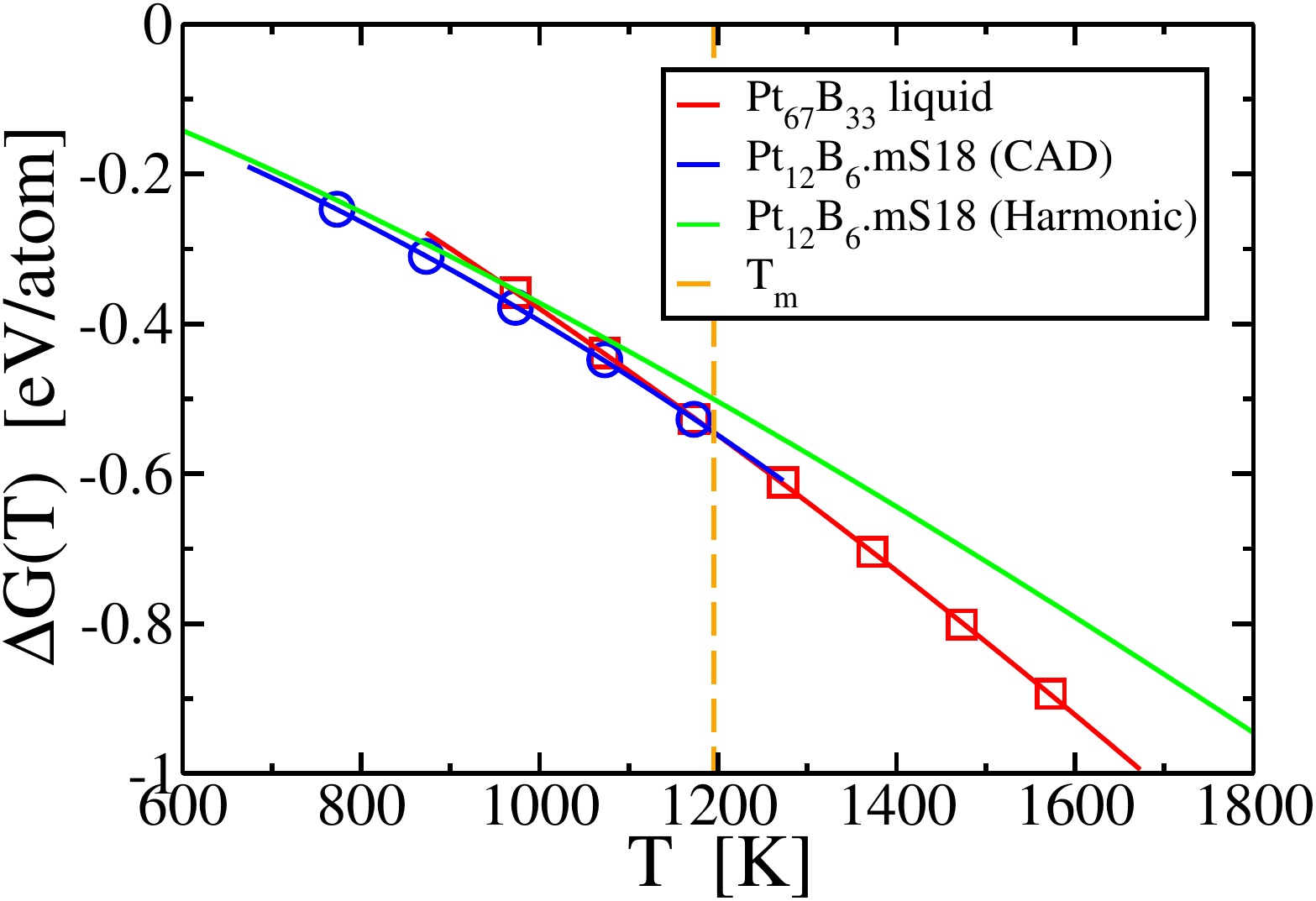}
\caption{\label{fig:G} Gibbs free energy as function of temperature for the harmonic solid (green), the anharmonic solid (blue) and the liquid state (red).}
\end{figure}

Temperature-dependent enthalpies $H(T)$ are obtained from the AIMD simulated total energies (including kinetic energy) at the temperature-dependent zero-pressure equilibrium volumes.  Fig.~\ref{fig:G} plots Gibbs free energies
\begin{equation}
  \label{dq:G}
  G(T)=H(T)-TS(T)
\end{equation}
for liquid and solid states. Data points show the results of calculations at the temperatures indicated, while the continuous curves represent fits of the data points to a quadratic polynomial, and these fits allow us to extrapolate beyond the temperature range covered by or simulations. For comparison with the the CAD method we also plot the harmonic free energy based on the phonon density of states of the Pt$_{12}$B$_6$ structure relaxed at $T=0$K. The relaxed energy is taken as the reference point for all three $G(T)$ curves.

Observe that the harmonic free energy crosses the liquid free energy at 957~K, which lies below the experimentally reported maximum solid phase temperature of 890~C (1163~K). In contrast, the anharmonic free energy obtained from CAD lies strictly below the harmonic free energy, and crosses the liquid free energy at $T_m=1195$~K, close to experiment. The predicted enthalpy and entropy of fusion are $\Delta H(T_m)=147$~meV/atom and $\Delta S(T_m)=1.45~\kB$. Note that 1195~K lies above our highest simulated temperature for the solid; we were unable to apply the CAD method at T=1273~K owing to the onset of atomic diffusion.

\section{Discussion}
\label{sec:discussion}

Metallurgical phase diagram investigations often match imprecisely with subsequent crystallographic structure refinements. The difficulty is enhanced for metal borides because strong covalent bonding may inhibit equilibration, and because boron atoms are difficult to localize by diffraction experiments. Reconciling phase diagrams as assessed by the ASM with crystallographic structures as reported by the ICSD is an important task that can be aided by first principles total energy investigation to help identify valid structures and their compositions. Our study of the B-Pt alloy system reveals that the hP4 structure of PtB is not an equilibrium phase, that the phase named Pt$_3$B$_2$ is actually Pt$_4$B$_x$.oS8, that Pt$_{12}$B$_6$.mS18 is highly stable, that modulated supercells of Pt$_3$B.hP6 are potentially stable at low T, and that Pt$_4$B.cI38 is likely stabilized by impurities. Many other metal boride phase diagrams could benefit from similar studies.

A deep eutectic requires either an unusually high liquid state entropy or low enthalpy, relative to the competing solid phases, or a combination of the two. In the case of B-Pt, aspects of each are present. Firstly, the enthalpy of fusion is quite low, considering the T=0K formation enthalpy of the crystalline solid, indicating an energetic contribution to the stability of the liquid. Indeed, we calculated the enthalpies of solid B (836 atoms in a $2\times 2\times 2$ rhombohedral supercell) and Pt (500 atoms in a $5\times 5\times 5$ cubic supercell), and liquid Pt$_{67}$B$_{33}$ at 1173K and find the formation enthalpy of the liquid alloy at this temperature to be $\Delta H_{\rm For} = -103.8$meV/atom. The liquid achieves its low energy by surrounding the boron atoms with platinum in configurations that resemble the crystalline solid. Specifically, the majority of liquid state boron atoms are surrounded by six platinum atoms in configurations that resemble the crystalline solid. This allows for strong covalent B-Pt bonding even in the liquid state.

\begin{figure*}[htb]
\centering
\includegraphics[width=.27\textwidth]{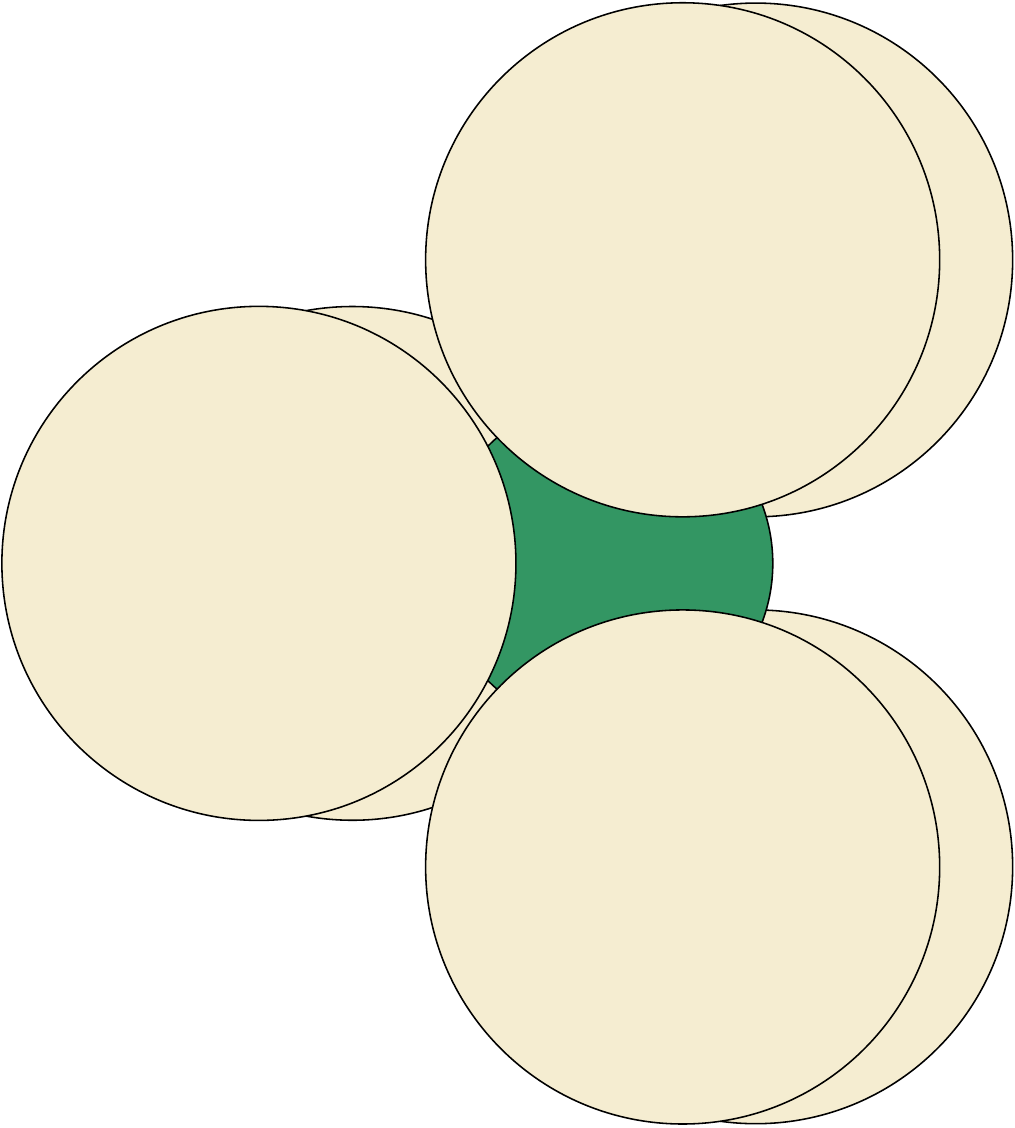}
\includegraphics[width=.3\textwidth]{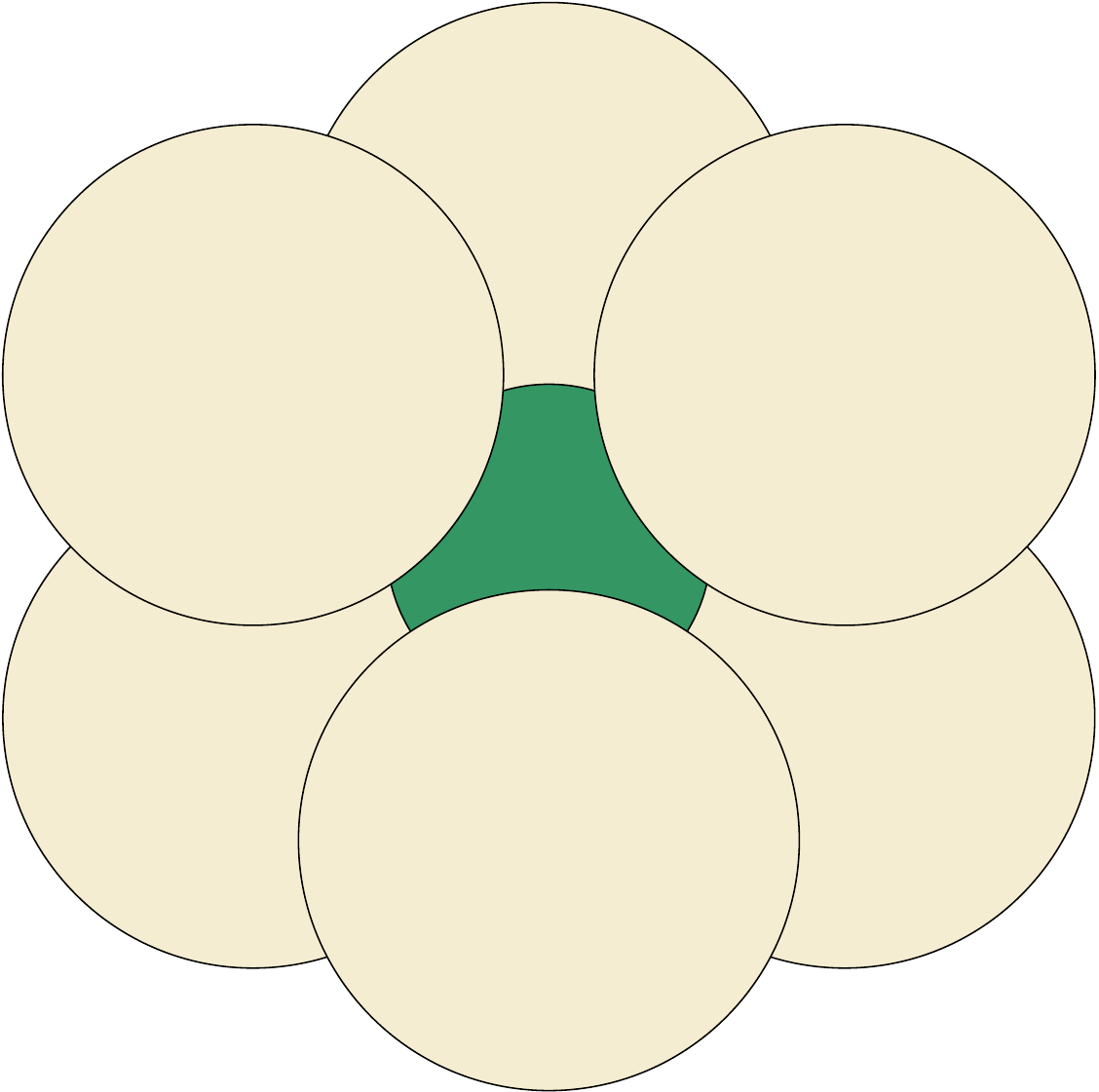}
\includegraphics[width=.3\textwidth]{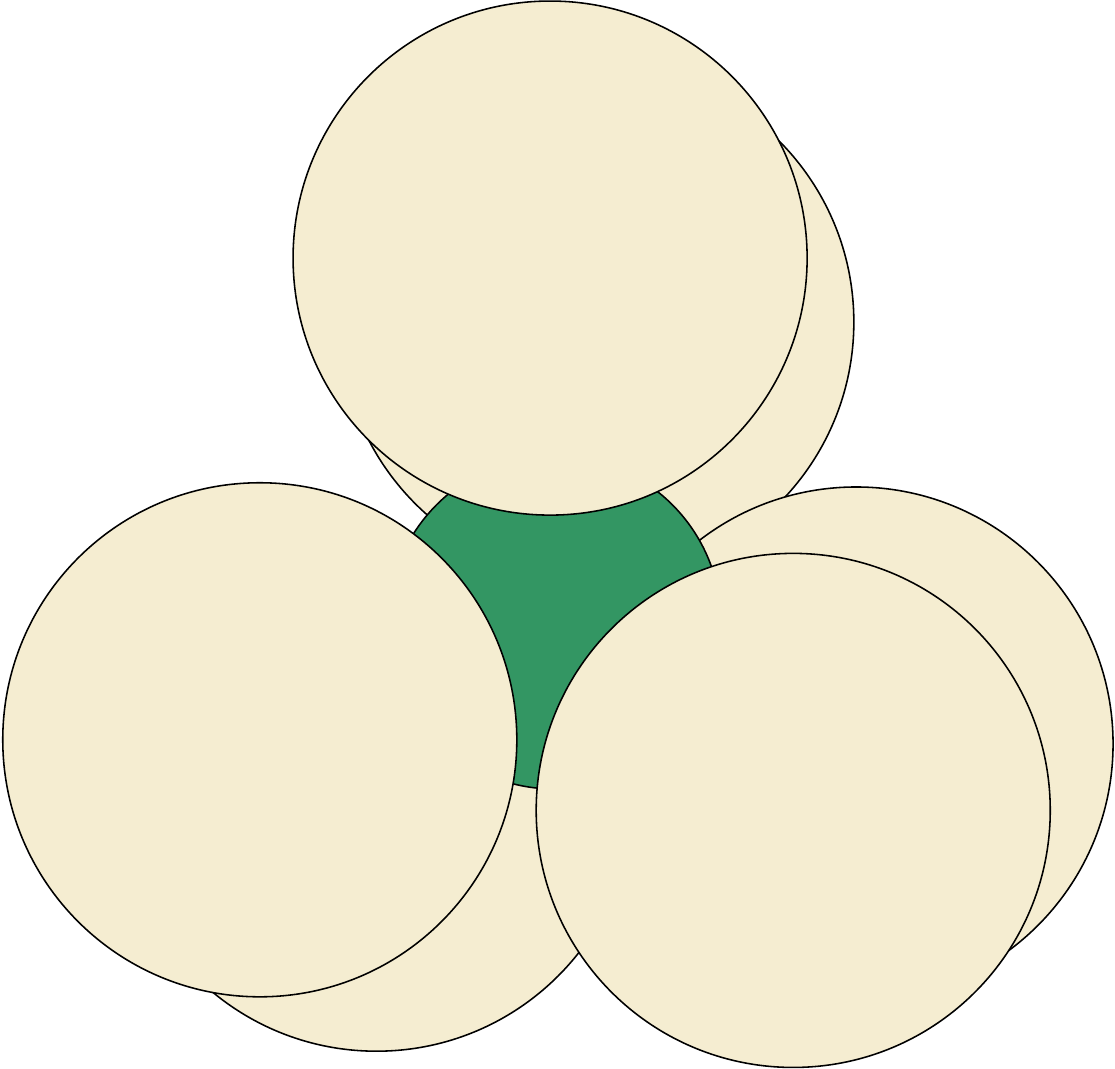}
\caption{\label{fig:XBs} Local boron environments in Pt$_2$B. (a) trigonal prism~\cite{fezryb} at the 4i site of .mS18; (b) trigonal anti-prism (octahedron) at the 2b site of mS18; (c) distorted trigonal prism in our simulated liquid at 1173K.}
\end{figure*}

Secondly, the entropy of the solid is not large. We calculated the entropy of elemental platinum at $T=1173$K to be $9.7123~\kB$/atom . Owing to partial site occupation and to the atomic diffusion in elemental boron, our CAD method is not capable of predicting its entropy. Fortunately, the value $S=33.069$~J/K/mol ($3.977~\kB$/atom) can be found in the JANAF Tables~\cite{JANAFB001}. Note that our calculated entropy of $8.513~\kB$ for the solid alloy at 1173K exceeds the composition-weighted mean of elemental B and Pt by $0.712~\kB$/atom, while the liquid entropy of $10.05~\kB$/atom exceeds it by three times that amount.

\section{Acknowledgements}
This work was supported by the Department of Energy under Grant No. DE-SC0014506. This research also used the resources of the National Energy Research Scientific Computing Center (NERSC), a US Department of Energy Office of Science User Facility operated under contract number DE-AC02-05CH11231. We thank Prof. Peter Rogl for numerous useful discussions.

\appendix
\section{Ab-initio molecular dynamics}
\label{app:AIMD}

We take simulation cells of 480 atoms for the liquid and 360 atoms for the solid, with a single electronic $k$-point but an elevated plane wave energy cutoff of 400 eV to minimize the Pulay stress. The NPT ensemble is applied initially to obtain thermal expansion at zero pressure, followed by NVT simulations for energy and entropy calculation. The inconsistent ensembles lead to small systematic errors in finite size simulation cells. A cubic volume constraint was applied for NPT simulations of the liquid owing to its vanishing shear modulus, but was not required for NPT simulations of the solid. Our NPT runs all exceeded 10ps, not including initial equilibration runs.

Simulated lattice parameters were averaged over the AIMD run then fit to quadratic functions in order to predict smoothly varying thermal expansion. Liquid state volumes fit to $V = 13.223 + 9.0679\times 10^{-4}~T$ in units of ~\AA$^3$/atom over the range 973-1573K. Solid state lattice constants fit to:
$a = 16.445 + 1.08\times 10^{-4}~T$;
$b = 16.332 + 2.62\times 10^{-4}~T$;
$c = 17.948 + 3.42\times 10^{-4}~T$ in units of ~\AA~ over the range 773-1173K.
The monoclinic angle $\gamma$ was nearly constant at the value 104.4$^\circ$. These smoothed values were then employed in our NVT simulations, which ran for 20ps each following the initial equilibration.

\begin{figure}[t!]
\centering
\includegraphics[width=.49\textwidth]{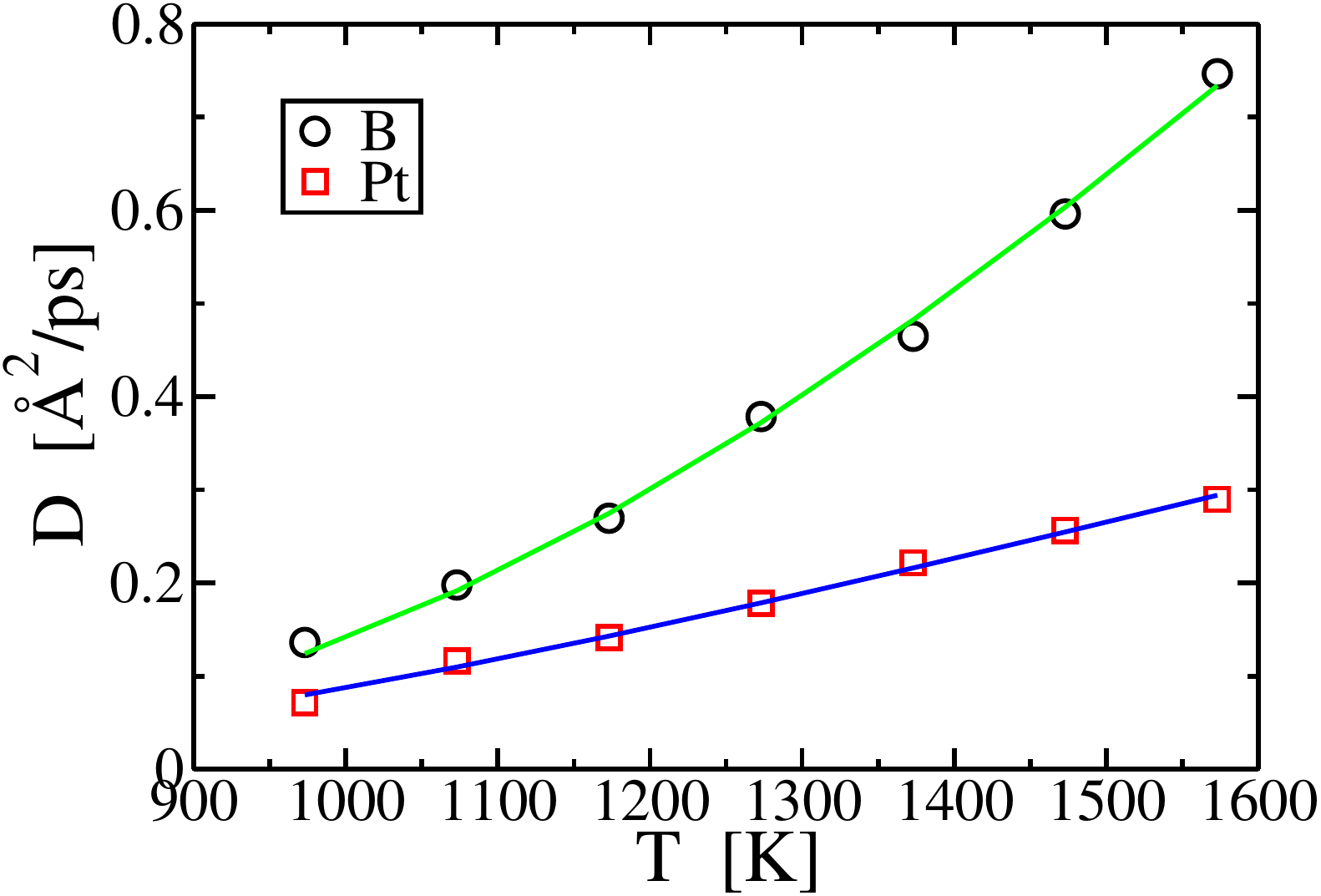}
\caption{\label{fig:Diff} Temperature-dependent diffusion constants predicted by AIMD.}
\end{figure}

Liquid state diffusion constants are available as a byproduct of our simulations. We extracted the diffusion constants shown in Fig.~\ref{fig:Diff} by fitting mean-square displacement to the form $\langle |{\bf r}|^2\rangle = d_0+6Dt$ over the time range 2.5 to 7.5 ps. The diffusion constant of boron fits to a clear Arrhenius form $D=A\exp{(-E_0/\kB T)}$, with an activation energy of $E_0=0.39$~eV, consistent with covalent bonding of B to its Pt neighbors. The fitted activation energy for Pt is 0.29 eV.

Machine learning features of VASP version 6.4.1 accelerated our simulations~\cite{VASP-ML}. A typical self-consistent AIMD time step for our 480 atom Pt$_{67}$B$_{33}$ liquid required approximately 800 seconds of CPU time on a 40 core Intel XEON processor, compared with 0.5 seconds to complete one ML time step, for a theoretical speed up of 1600$\times$. However, we chose to run in training mode in order to achieve long run times while limiting the errors inherent in the ML force fields. VASP automatically  reverts to full {\em ab-initio} forces and energies whenever the Bayesian error estimates exceed a target. In the liquid state NPT simulations 3-8\% of steps were performed ab-initio, limiting the actual speedups to the range of 10-20$\times$. In the solid state NVT simulations our effective speedup reached up to 50$\times$. During the NPT runs the RMS regression errors for liquids were 1.4 meV/atom in energy, 0.13 eV/\AA~ in force components, and 0.7 kbar in stress components. The RMS regression errors for solids were 0.6 meV/atom in energy, 0.09 eV/\AA~ in force, and 0.8 kbar in stress components.









\end{document}